\documentclass{ws-procs9x6}
\usepackage{amsmath,amssymb}

\newif\ifuseprd
\useprdfalse
\newif\ifeprint
\eprinttrue
\newif\ifdatelast
\datelasttrue
\newif\iftoomuchdetail
\toomuchdetailfalse
\let\oldappendix\appendix
\renewcommand\appendix{\oldappendix%
    \renewcommand\theequation{\thesection.\arabic{equation}}}
\newcounter{saveequation}
\newcounter{detailnum}
\newcommand\savetheequation{\theequation}
\newcommand\detailtheequation{%
	  $\delta$\Roman{detailnum}:\roman{equation}}

\newenvironment{detail}{\iftoomuchdetail\sf% no let's set up equations
         \setcounter{saveequation}{\value{equation}}%
         \setcounter{equation}{0}\addtocounter{detailnum}{1}%
         \renewcommand\theequation\detailtheequation%
         \fi}{% unset up equations
     \iftoomuchdetail%
     \ifnum\value{equation}=0\addtocounter{detailnum}{-1}\fi%
     \setcounter{equation}{\value{saveequation}}%
     \renewcommand\theequation\savetheequation%
     \fi%
     }

\makeatletter
\def\@strike{\relax\leavevmode
  \ifmmode
    \expandafter\mathpalette\expandafter\math@strike
  \else
    \expandafter\make@strike
  \fi}
\def\math@strike#1#2{%
  \setbox\z@\hbox{$\m@th#1{#2}$}\fin@strike}
\def\make@strike#1{%
  \setbox\z@\hbox{\color@begingroup#1\color@endgroup}\fin@strike}
\def\fin@strike{%
  \@tempdima\dp\z@
  \@tempdimb\ht\z@
  \lower\@tempdima\hbox{\strike@start}%
  \box\z@
  \raise\@tempdimb\hbox{\strike@end}}
\def\strike@start{\special{ps: %
    currentpoint /starty exch def /startx exch def}}
\def\strike@end{
}
\newcommand\fs{\protect\@strike}

\newcommand\p{\ensuremath{\partial}}
\newcommand\evalat[2]{\ensuremath{\left.{#1}\right|_{#2}}}

\newcommand\transpose{{\ensuremath{\text{\sf T}}}}
\newcommand\field[1]{{\ensuremath{\mathbb{{#1}}}}}

\newcommand\ZR{{\field{R}}}

\newcommand\anti[2]{\ensuremath{\left\{{#1},{#2}\right\}}}
\newcommand\com[2]{\ensuremath{\left[{#1},{#2}\right]}}
\DeclareMathOperator{\Tr}{Tr}

\newcommand\mathone{{\rlap{\kern .25em l}1}}
\newcommand\one{{\ifmmode{\text{\mathone}}\else{\mathone}\fi}}
\newcommand\proj{{\ensuremath{{{\mathbb P}}}}}
\newcommand\apr{{\ensuremath{{\alpha'}}}}

\newcommand\lie[2]{\ensuremath{\pounds_{{#1}} {#2}}}
\providecommand\FIGURE[2][]{\begin{figure}[#1]\begin{center}{#2}\end{center}
                       \end{figure}}

\newcommand\citejournal[4]{{\ifuseprd\else\begingroup\em\fi {#4}%
     \ifuseprd\else\endgroup\fi {\bf {#1}}\ifdatelast, {#3} ({#2})\else%
     \ ({#2}) {#3}\fi}}

\providecommand\plb[3]{{\citejournal{B#1}{#2}{#3}{Phys.\ Lett.\ }}}
\providecommand\npb[3]{{\citejournal{B#1}{#2}{#3}{Nucl.\ Phys.\ }}}
\providecommand\jhep[3]{{\citejournal{#1}{#2}{#3}{J.\ High Energy Phys.\ }}}

\providecommand\cqg[3]{{\citejournal{#1}{#2}{#3}{Class.\ Quant.\ Grav.\ }}}

\providecommand\cmp[3]{{\citejournal{#1}{#2}{#3}{Commun.\ Math.\ Phys.\ }}}

\providecommand\citeprd[3]{{\citejournal{D#1}{#2}{#3}{Phys.\ Rev.\ }}}

\providecommand\hepth[1]{{\ifuseprd{\eprint{{\ifeprint\tt\fi hep-th/#1}}}%
                \else{\tt hep-th/{#1}}\fi}}

\newcommand\phepth[1]{{\ifuseprd\else\tt\fi [\hepth{#1}]}}
\newcommand\ct[1]{{\ifeprint\ifuseprd{\em{#1}},\else{``{#1},''}\fi%
      \else$\!\!$\fi}}
\newcommand\bt[1]{{\em {#1}},}

\newcommand\skipthis[1]{{}}

\newcommand\ang{{\mathcal J}}
\newcommand\notang{{\mathcal K}}

\newcommand\ppwave{{\em pp}~wave}
\newcommand\ppWave{{\em pp}~Wave}
\newcommand\supersymmetry{SUSY}
\newcommand\supersymmetries{SUSYs}

\newcommand\insertdualityfig{\FIGURE[t]{
\setlength{\unitlength}{\baselineskip} % so everything is in units of
\begin{picture}(22,29)(-6,6) % made it too long and narrow at first, so offset
\put(1,31){\frame{
   \begin{picture}(8,4)
      \put(1.5,1){\shortstack{
         {\bf M-Theory:} \\ \\
         $R_-, \quad R_8, \quad R_9$ \\
         $\ell_P$}}
   \end{picture}}}
\put(10,23){\frame{
   \begin{picture}(6,6)
      \put(1,1){\shortstack{
         {\bf IIA:} \\ \\
         $R_8, \quad R_9$ \\
         $g_s = \left( \frac{R_-}{\ell_P} \right)^{3/2}$ \\
         $\ell_s = \left( \frac{\ell_P^3}{R_-} \right)^{1/2}$}}
   \end{picture}}}
\put(7,31){\vector(3,-1){6}} % that is, slope of y/x and horizontal
   \put(11.5,30){\makebox(0,0){on $-$}}
\put(10,15){\frame{
   \begin{picture}(6,6)
      \put(1,1){\shortstack{
         {\bf IIB:} \\ \\
         $\frac{\ell_P^3}{R_8 R_-}, \quad R_9$ \\
         $g_s = \frac{R_-}{R_8}$ \\
         $\ell_s = \left( \frac{\ell_P^3}{R_-} \right)^{1/2}$}}
   \end{picture}}}
\put(13,23){\vector(0,-1){2}} \put(13,21){\vector(0,1){2}}
   \put(14,22){\makebox(0,0){$T8$}}
\put(10,6){\linethickness{.2em}\frame{
   \begin{picture}(6,7)
      \put(1,1){\shortstack{
         {\bf IIA:} \\ \\
         $\frac{\ell_P^3}{R_8 R_-}, \quad \frac{\ell_P^3}{R_9 R_-}$ \\
         $g_s = \frac{\ell_P^{3/2} R_-^{1/2}}{R_8 R_9}$ \\
         $\ell_s = \left( \frac{\ell_P^3}{R_-} \right)^{1/2}$}}
   \end{picture}}}
\put(13,15){\vector(0,-1){2}} \put(13,13){\vector(0,1){2}}
   \put(14,14){\makebox(0,0){$T9$}}
\put(2,23){\frame{
   \begin{picture}(6,6)
      \put(1,1){\shortstack{
         {\bf IIA:} \\ \\
         $R_-, \quad R_8$ \\
         $g_s = \left( \frac{R_9}{\ell_P} \right)^{3/2}$ \\
         $\ell_s = \left( \frac{\ell_P^3}{R_9} \right)^{1/2}$}}
   \end{picture}}}
\put(5,31){\vector(0,-1){2}}
   \put(4.5,30){\makebox(0,0)[r]{on $9$}}
\put(8,26){\vector(1,0){2}} \put(10,26){\vector(-1,0){2}}
   \put(9,26){\makebox(0,0){\shortstack{9/$-$ \\ flip}}}
\put(2,15){\linethickness{.2em}
  \frame{
   \begin{picture}(6,6)
      \put(1,1){\shortstack{
         {\bf IIB:} \\ \\
         $R_-, \quad \frac{\ell_P^3}{R_8 R_9}$ \\
         $g_s = \frac{R_9}{R_8}$ \\
         $\ell_s = \left( \frac{\ell_P^3}{R_9} \right)^{1/2}$}}
   \end{picture}}}
\put(5,23){\vector(0,-1){2}} \put(5,21){\vector(0,1){2}}
   \put(4,22){\makebox(0,0){$T8$}}
\put(2,6){\frame{
   \begin{picture}(6,7)
      \put(1,1){\shortstack{
         {\bf IIA:} \\ \\
         $\frac{\ell_P^3}{R_9 R_-}, \quad \frac{\ell_P^3}{R_8 R_9}$ \\
         $g_s = \frac{\ell_P^{3/2} R_9^{1/2}}{R_8 R_-}$ \\
         $\ell_s = \left( \frac{\ell_P^3}{R_9} \right)^{1/2}$}}
   \end{picture}}}
\put(5,15){\vector(0,-1){2}} \put(5,13){\vector(0,1){2}}
   \put(4,14){\makebox(0,0){$T-$}}
\put(8,9.5){\vector(1,0){2}} \put(10,9.5){\vector(-1,0){2}}
   \put(9,9.5){\makebox(0,0){\shortstack{9/$-$\\ flip}}}
\put(-6,23){\frame{
   \begin{picture}(6,6)
      \put(1,1){\shortstack{
         {\bf IIA:} \\ \\
         $R_-, \quad R_9$ \\
         $g_s = \left( \frac{R_8}{\ell_P} \right)^{3/2}$ \\
         $\ell_s = \left( \frac{\ell_P^3}{R_8} \right)^{1/2}$}}
   \end{picture}}}
\put(3,31){\vector(-3,-1){6}} % that is, slope of y/x and horizontal
   \put(-1,30){\makebox(0,0)[r]{on $8$}}
\put(0,26){\vector(1,0){2}} \put(2,26){\vector(-1,0){2}}
   \put(1,26){\makebox(0,0){\shortstack{8/9\\ flip}}}
\put(-3,23){\vector(0,-1){2}} \put(-3,21){\vector(0,1){2}}
   \put(-4,22){\makebox(0,0)[r]{$T9$}}
\put(-6,15){%\linethickness{.2em}
  \frame{
   \begin{picture}(6,6)
      \put(1,1){\shortstack{
         {\bf IIB:} \\ \\
         $R_-, \quad \frac{\ell_P^3}{R_8 R_9}$ \\
         $g_s = \frac{R_8}{R_9}$ \\
         $\ell_s = \left( \frac{\ell_P^3}{R_8} \right)^{1/2}$}}
   \end{picture}}}
\put(0,18){\vector(1,0){2}} \put(2,18){\vector(-1,0){2}}
   \put(1,19){\makebox(0,0){$S$}}
\put(-6,6){\frame{
   \begin{picture}(6,7)
      \put(1,1){\shortstack{
         {\bf IIA:} \\ \\
         $\frac{\ell_P^3}{R_8 R_-}, \quad \frac{\ell_P^3}{R_8 R_9}$ \\
         $g_s = \frac{\ell_P^{3/2} R_8^{1/2}}{R_9 R_-}$ \\
         $\ell_s = \left( \frac{\ell_P^3}{R_8} \right)^{1/2}$}}
   \end{picture}}}
\put(-3,15){\vector(0,-1){2}} \put(-3,13){\vector(0,1){2}}
   \put(-4,14){\makebox(0,0){$T-$}}
\put(0,9.5){\vector(1,0){2}} \put(2,9.5){\vector(-1,0){2}}
   \put(1,9.5){\makebox(0,0){\shortstack{8/9\\ flip}}}
\end{picture}
\caption{The type IIB maximally supersymmetric \ppwave, compactified
on $x^8$ (and with a compact null circle), is described by a IIA 2+1
SYM (D2-brane worldvolume theory) via a series of dualities from
M-theory.  In each box, the
first line specifies the compactification radii.  The IIB theory of
interest, and the IIA theory of D2-branes which describes it, are
highlighted.\label{fig:duality}}
}
\renewcommand\insertdualityfig{} % so don't use it twice!
}

\ifeprint\def\draftnote{}\fi

\begin{document}

%%%%%%%%%%%%%%%%%%%%%%%%%%%%%%%%%%%%%%%%%%%%%%%%%%%%%%%%%%%%%% 
% title, author(s) and address(es) put here:                 %
%%%%%%%%%%%%%%%%%%%%%%%%%%%%%%%%%%%%%%%%%%%%%%%%%%%%%%%%%%%%%% 

\title{
\ifeprint
\vspace{-3\baselineskip}
\begingroup
\footnotesize\normalfont\raggedleft
\lowercase{\sf hep-th/0401050} \\ UK/04-01 \\
\vspace{\baselineskip}
\endgroup
\fi
Matrix Theory of {\em \lowercase{pp}\/} Waves%
\ifeprint
\footnote{%{\uppercase{P}}resented \uppercase{S}ept.\ 12, 2003 at 
\uppercase{T}o appear in the \uppercase{P}roceedings of
the 3rd 
\uppercase{I}nternational \uppercase{S}ymposium on \uppercase{Q}uantum
\uppercase{T}heory and \uppercase{S}ymmetries (\uppercase{QTS}3) 
%\uppercase{C}incinnati, \uppercase{O}hio.
10~--~14 \uppercase{S}ept 2003 ---
%\uppercase{P}roceedings
\copyright\
\uppercase{W}orld \uppercase{S}cientific.}
% Who thought footnotes should be lowercase only???
\fi
}

\author{Jeremy Michelson}
\address{Department of Physics and Astronomy \\
 University of Kentucky \\
 Lexington, KY~~40506~~USA \\
 E-mail: jeremy@pa.uky.edu}  


\maketitle

\abstracts{
The Matrix
Theory that has been proposed for various \ppwave\ backgrounds
is discussed.  Particular emphasis is on
the existence of novel nontrivial supersymmetric solutions of the Matrix
Theory. These correspond to branes of various shapes
(ellipsoidal, paraboloidal, and possibly hyperboloidal) that are
unexpected from previous studies of branes in \ppwave\ geometries.
\iftoomuchdetail
%begin{detail}%
\begingroup\sf
A Matrix String theory for the maximally supersymmetric type IIB \ppwave\
is also discussed.
\endgroup%
%\end{detail}%
\fi
}

%%%%%%%%%%%%%%%%%%%%%%%%%%%%%%%%%%%%%%%%%%%%%%%%%%%%%%%%%%%%%
% The main text of your paper                               %
%%%%%%%%%%%%%%%%%%%%%%%%%%%%%%%%%%%%%%%%%%%%%%%%%%%%%%%%%%%%%

\section{INTRODUCTION}

The BMN\cite{bmn} Matrix Theory, which describes the maximally
supersymmetric \ppwave\ of M-theory\cite{kg}, is a very nice arena
for learning about M-theory.  It is simple enough to be reasonably
tractable, yet describes a curved background.  Moreover, it has a
dimensionless parameter $\frac{\mu \apr}{R}$---the ratio, in string
units, between the strength of the four-form flux and the DLCQ
radius---which, when large, permits a perturbative treatment of the
Yang-Mills theory\cite{dsv1}.  Additionally, the four-form flux
supports fuzzy sphere solutions to the Matrix Theory\cite{bmn}.  The
fuzzy spheres are the general vacua of the Matrix Theory, which
preserve {\em all\/} the \supersymmetries.  These fuzzy spheres have
very interesting properties upon compactification\cite{jm} of the
theory to a IIA Matrix String Theory.  These results have already
appeared in the literature\cite{dms,cs} and space constraints prevent
us from summarizing them here.

There are many \ppwave\ solutions to
M-theory\cite{clp,clp2,ghpp,jm2} and a Matrix Theory has
been proposed\cite{clp2,ni} for each.
The original argument noted that the form
reproduces BMN and has the right number of \supersymmetries\cite{clp2}.
A derivation from membrane quantization has since been given\cite{ni}.
One can also show,
beyond
counting fermionic generators, that
the full \supersymmetry\ algebras match\cite{jmno}.

Having proposed a Matrix Theory for each M-theory \ppwave, one should
now study them.  The zeroth question to answer is: what are the vacua
which preserve all of the \supersymmetries?  We will study this for
three \ppwave{s}: the Penrose
limit of AdS$_3\times S^3$ \cite{clp,clp2,rt,gms}; the T-dual\cite{jm} of the
maximally supersymmetric \ppwave\ of the IIB theory\cite{bfhp}; and the
26 supercharge \ppwave\cite{jm2}.  For the first two, we find, at
infinite $N$, brane solutions which do not follow from treating
$g_{++}$ as a superpotential\cite{hy}.%
\footnote{%All papers on branes in \ppwave\ backgrounds are too
%numerous to list here.
%I wish there was space 
Unfortunately there is no room
to list the myriad papers on
branes in \ppwave{s}.}

\iftoomuchdetail
\begin{detail}%
Since we will discuss the T-dual of
the IIB \ppwave, we should also attempt to make contact with IIB
physics.  This is the subject of \S\ref{sec:2+1}.
\end{detail}%
\fi

\section{MATRIX THEORIES} \label{sec:mt}

The general 11-dimensional \ppwave\ %(M\ppwave)
considered is\cite{fp,clp2,ghpp}
\begin{equation} \label{Mpp}
%\begin{aligned}
ds^2 = 2 dx^+ dx^- - \left[ \sum_{i=1}^9 \mu_i^2 (x^i)^2 \right] (dx^+)^2
   + (dx^i)^2,
\qquad
F = dx^+ \wedge \Theta,
%\end{aligned}
\end{equation}
with constant $\mu_i$, $\Theta$,
and also $\Theta$ is a ``spatial'' (support in the $x^i$-directions)%
\footnote{The index $i,j,\dots$ runs from 1 to 9, and $A,B,\dots$ runs
over all coordinates.  $\Gamma$-matrices are defined with respect to
the elfbein
\hbox{$e^{\hat{-}} = dx^- - \frac{1}{2} \sum_{i=1}^9 \mu_i^2 (x^i)^2 dx^+$},
\hbox{$e^{\hat{+}} = dx^+$},
\hbox{$e^{\hat{\imath}} = dx^i$}.  The Feynman slash, {\em e.g.\/}\ 
\hbox{$\fs{\Theta} = \frac{1}{3!} \Gamma^{ijk} \Theta_{ijk}$}, is used
extensively.}
three-form, satisfying the equation of motion (e.o.m.)
\hbox{$\Theta_{ijk}^2 = 12 \sum_i
\mu_i^2$}.

This geometry admits $16+2n$ Killing spinors.  To present them---and
define $n$---set, (with all sums
explicit here, and 
\hbox{$U_{(i)} \equiv 3 \Gamma^i \fs{\Theta} \Gamma^i + \fs{\Theta}$})
\begin{gather}
\begin{aligned}
\Omega_+ &\equiv -\frac{1}{12} \fs{\Theta}(\Gamma^+ \Gamma^- + \one), &
\Omega_- &\equiv 0, &
\Omega_i &\equiv \frac{1}{24} \Gamma^i U_{(i)} \Gamma^+.
\end{aligned} \label{defomega}
\end{gather}
The solutions of the Killing spinor equation,
%\begin{equation} \label{kseq}
\hbox{${\mathcal D}_A \epsilon = \nabla_A \epsilon - \Omega_A \epsilon = 0$},
%\end{equation}
are
\begin{equation} \label{MppK}
%\hbox{$
\epsilon(x,\epsilon_0) = \Bigl[1+\sum_i x^i \Omega_i\Bigr]
   e^{-\frac{1}{12} (\Gamma^+ \Gamma^-+\one)\fs{\theta} x^+} \epsilon_0,
%$}
\end{equation}
where the constant spinor $\epsilon_0$ obeys\cite{fp,clp2,ghpp}
\begin{equation}
U_{(i)}^2 \Gamma^+ \epsilon_0 = -144 \mu_i^2 \Gamma^+ \epsilon_0.
\end{equation}
So there are 16 ``standard'' \supersymmetries\ which obey $\Gamma^+
\epsilon_0=0$ and an even number, $2n$, of
``supernumerary''\cite{clp} \supersymmetries\ obeying
\begin{equation} \label{susymu}
U_{(i)}^2 \epsilon_0 = -144 \mu_i^2 \epsilon_0,
\end{equation}
{\em for each $i$\/}.  Equation~\eqref{susymu} need not have
any solutions, and never has 12 or 14 solutions\cite{jm2}.
Existence of ``supernumerary'' \supersymmetries\
guarantees a solution to the e.o.m.s, by
%The existence of ``supernumerary'' \supersymmetries
%implies the equation of motion, as follows from
summing
Eq.~\eqref{susymu} over $i$ and tracing over %the subspace of
Killing spinors.
\iftoomuchdetail
\begin{detail}%
Explicitly, if $\proj$ projects onto the subspace of
solutions to Eq.~\eqref{susymu}, then Eq.~\eqref{susymu}
implies that
\begin{equation} \label{susyeom}
\sum_i \Tr U_{(i)}^2 \proj =
-12 \Theta_{ijk} \Theta^{ijk} \Tr \proj = -144 \sum_i \mu_i^2 \Tr \proj,
\end{equation}
using $\sum_i \Gamma^i \fs{\Theta} \Gamma^i = -3 \fs{\Theta}$.
So if the e.o.m.\ is not satisfied then $\Tr \proj=0$.
\end{detail}%
\fi

In units for which the 11-dimensional Planck length, $\ell_P=1$,
and employing Majorana
${\mathfrak so}(9)$ spinors $\Psi$,
the Matrix
Theory action\cite{clp2,ni} is%
\footnote{\iftoomuchdetail
\begin{detail}%
Conventions include
\end{detail}%
\fi
$D_\tau X^i = \p_\tau X^i + i \com{A_\tau}{X^i}$ and
\iftoomuchdetail
\begin{detail}%
the fact that complex conjugation reverses
the order of Grassmann variables,
\end{detail}%
\fi
$(\Psi_\alpha \Psi_\beta)^* =
\Psi^*_\beta \Psi^*_\alpha$.
$R$ is the DLCQ radius.}
\begin{multline} \label{MppQM}
S = R \int d\tau \Tr \left\{ \frac{1}{2 R^2} (D_\tau X^i)^2
  + \frac{i}{R} \Psi^\transpose D_{\tau} \Psi
  + \Psi^\transpose \Gamma^i \com{X^i} \Psi
  + \frac{1}{4} \com{X^i}{X^j}^2 
\right. \\ \left.
-\frac{1}{2} \sum_i \frac{\mu_i^2}{R^2} (X^i)^2
-\frac{i}{4 R} \Psi^\transpose \fs{\Theta} \Psi
-\frac{i}{3 R} \Theta_{ijk} X^i X^j X^k
\right\}.
\end{multline}
%As for BMN\cite{bmn}, t
The first line is the familiar BFSS Matrix Theory\cite{bfss}
action; the second line
contains mass terms for the bosons and fermions% from the \ppwave
, and
the Myers term\cite{rm,tv}.

The Matrix Theory~\eqref{MppQM} is invariant under the
nonlinearly realized \supersymmetry\ 
%\begin{equation} \label{nlSUSY}
\hbox{$
\delta \Psi = e^{\frac{1}{4} \fs{\Theta} \tau} \epsilon_+%,
$}
%\end{equation}
%where $\epsilon_+$ is an arbitrary constant spinor.
corresponding
to the 16 ``standard'' \supersymmetries\ of the \ppwave.
\iftoomuchdetail%GAMMA APPENDIX
\begin{detail}%
The sign in the exponential follows from Eq.~\eqref{minussign}.
\end{detail}%
\fi
Consider also the transformations,
\begin{subequations} \label{MppSUSY}
\begin{gather} \label{MppSUSYX}
\begin{align} 
\delta X^i &= i \Psi^\transpose \Gamma^i \epsilon(\tau), &
\delta A_\tau &= i R \Psi^\transpose \epsilon(\tau), &
\epsilon(\tau) & \equiv e^{-\frac{1}{12} \fs{\Theta} \tau} \epsilon_0,
\end{align} \\ \label{MppSUSYPsi}
\delta \Psi = \frac{1}{2R} D_\tau X^i \Gamma^i \epsilon(\tau)
 + \frac{1}{R} X^i \hat{\Omega}_i \epsilon(\tau)
 + \frac{i}{4} \com{X^i}{X^j} \Gamma^{ij} \epsilon(\tau),
\end{gather}
\end{subequations}
where $\hat{\Omega}_i \equiv \frac{1}{24} \Gamma^i U_{(i)}$ and
$\epsilon_0$ is constant%.
%Similar expressions have been given by 
\ ({\em cf.\/}\ Bonelli\cite{gb}).
These transformations preserve the
action precisely when $\epsilon_0$ obeys%
\iftoomuchdetail
\begin{detail}%
\ the ``supernumerary''
\supersymmetry\ condition%
\end{detail}%
\fi
~\eqref{susymu}.  So the Matrix quantum
mechanics~\eqref{MppQM} preserves exactly the right number of
\supersymmetries\ to describe the \ppwave~\eqref{Mpp}\cite{clp,ni}.
Indeed, the \supersymmetry\ algebras match\cite{jmno}.
\iftoomuchdetail
\begin{detail}%
The BMN model\cite{bmn} is a special case of~\eqref{MppQM}.
\end{detail}%
\fi

\section{THE AdS$_3\times S^3$ {\em pp\/} WAVE} \label{sec:AdS3}

The Penrose limit of $AdS_3\times S^3$ has been considered
in\cite{clp,clp2,rt,gms}.  Lifting the geometry supported with
NS-NS flux to 11-dimensions,
\begin{equation}
\begin{gathered}
ds^2 = 2 dx^+ dx^- - \mu^2 \sum_{a=1}^4 (x^a)^2 (dx^+)^2
    + \sum_{i=1}^9 (dx^i)^2, \\
^{(4)}F = 2\mu dx^+ \wedge dx^1 \wedge dx^2 \wedge dx^9
         +2\mu dx^+ \wedge dx^3 \wedge dx^4 \wedge dx^9.
\end{gathered}
\end{equation}
The M-theory direction is $x^9$, and the directions $x^a$,
$a{=}1\cdots4$ are the spacelike $AdS_3\times S^3$ directions transverse to the
null geodesic of the Penrose limit.

\iftoomuchdetail
\begin{detail}%
The Matrix Theory action~\eqref{MppQM} for this background is
\begin{multline} \label{AdS3ppQM}
S = R \int d\tau \Tr \left\{ \frac{1}{2 R^2} (D_\tau X^i)^2
  + \frac{i}{R} \Psi^\transpose D_{\tau} \Psi
  + \Psi^\transpose \Gamma^i \com{X^i}{\Psi}
  + \frac{1}{4} \com{X^i}{X^j}^2 
\right. \\ \left.
- \frac{\mu^2}{2 R^2} \sum_{a=1}^4 (X^a)^2
+ i\frac{\mu}{2 R} \Psi^\transpose (\Gamma^{129}+ \Gamma^{349}) \Psi
\right. \\ \left.
-2 i \frac{\mu}{R} \com{X^1}{X^2}{X^9}
-2 i \frac{\mu}{R}\com{X^3}{X^4}{X^9}
\right\}.
\end{multline}
The action is invariant under the 8 linearly realized
\supersymmetries\ generated by
\begin{subequations} \label{AdS3ppSUSY}
\begin{gather}
\begin{align}
\delta X^i &= i \Psi^\transpose \Gamma^i \epsilon_0, &
\delta A_\tau &= i R \Psi^\transpose \epsilon_0,
\end{align} \\
\begin{aligned} \label{AdS3dPsi}
\delta \Psi &= \frac{1}{2R} D_\tau X^i \Gamma^i \epsilon_0
 + \frac{\mu}{2R} X^1 \Gamma^{29} \epsilon_0
 - \frac{\mu}{2R} X^2 \Gamma^{19} \epsilon_0
\\ & \qquad
 + \frac{\mu}{2R} X^3 \Gamma^{49} \epsilon_0
 - \frac{\mu}{2R} X^4 \Gamma^{39} \epsilon_0
 + \frac{i}{4} \com{X^i}{X^j} \Gamma^{ij} \epsilon_0,
\end{aligned}
\intertext{where the constant spinor $\epsilon_0$ obeys} \label{AdS3eps}
\Gamma^{1234}\epsilon_0 = \epsilon_0.
\end{gather}
\end{subequations}
\end{detail}%
\fi
Vanishing of the fermionic variation~%
\iftoomuchdetail
\eqref{AdS3dPsi}
\else
\eqref{MppSUSYPsi}
\fi
gives the conditions
\begin{equation} \label{AdS3SUSYcond}%
\begin{gathered} %\label{12=34}
\begin{aligned}
\com{X^1}{X^2} &= \com{X^3}{X^4}, &
\com{X^2}{X^3} &= \com{X^1}{X^4}, &
\com{X^1}{X^3} &= -\com{X^2}{X^4},
\end{aligned} \\
\begin{aligned}
\com{X^2}{X^9} &= i \frac{\mu}{R} X^1, &
\com{X^1}{X^9} &= -i \frac{\mu}{R} X^2, \\
\com{X^4}{X^9} &= i \frac{\mu}{R} X^3, &
\com{X^3}{X^9} &= -i \frac{\mu}{R} X^4.
\end{aligned}
\end{gathered}
\end{equation}%
All other commutators vanish, and the vacua are static.

For finite $N$, all vacua preserving
all eight linearly-realized \supersymmetries\ are trivial:
\hbox{$X^1=X^2=X^3=X^4=0$}%
\iftoomuchdetail
\begin{detail} % need a space!
and the commutators of the massless directions
vanish.  This is easily seen by multiplying the leftmost equation
in~\eqref{AdS3SUSYcond} by $\com{X^1}{X^2}$, taking the trace, and using
cyclicity of the trace, the Jacobi identity and the other two
equations in~\eqref{AdS3SUSYcond} to obtain
\begin{equation}
\Tr \com{X^1}{X^2}^2 + \Tr \com{X^2}{X^3}^2 + \Tr \com{X^1}{X^3}^2 =
0,
\end{equation}
which implies that each commutator vanishes%
\end{detail}%
\fi
. However, there is a family of solutions for infinite $N$%
\iftoomuchdetail
\begin{detail}%
.\footnote{In section~\ref{sec:TIIB} we will find by, for example,
embedding the coordinates into semi-simple Lie algebras, supersymmetric vacua
of the \ppwave\ obtained from T-dualization of the IIB maximally
supersymmetric \ppwave.  There do not
appear to be any such vacua here, as can be seen, for example, by
noting that the two ``simple roots'' $X^1+i X^2$ and $X^3+i X^4$ have
equal, nonzero weights, which is a contradiction.  This does not mean
that the family of solutions we present are the only solutions of the
equations.}
\end{detail}%
\else
.
\fi
Specifically, take $X^1, X^2$ and $X^3, X^4$ to form two
noncommutative planes, with equal noncommutativity parameter,
\begin{equation}
\com{X^1}{X^2} = i \vartheta = \com{X^3}{X^4};
\qquad \com{X^1}{X^3} = 0 = \com{X^2}{X^3},
\end{equation}
and take%
\iftoomuchdetail
\begin{detail}
\footnote{Had we tried to break the SO(2)$\times$SO(2) symmetry of
the \ppwave\ by taking, say \hbox{$\com{X^2}{X^3}=i\vartheta=\com{X^1}{X^4}$}
with all other commutators in that fourspace vanishing, then 
for no choice of signs does \hbox{$X^9 =
\frac{\mu}{R \vartheta}[ \pm X^1 X^3 \pm' X^2 X^4 ]$} give correct
commutation relations.  Interestingly, this is the same conspiracy of
signs that prevents us finding solutions via an embedding into a
noncompact form of ${\mathfrak so}(4)$.}
\end{detail}
\fi
\begin{equation} \label{parab}
X^9 = X^9_0 -\frac{\mu}{2 R \vartheta} \left[
  (X^1)^2 + (X^2)^2 + (X^3)^2 + (X^4)^2 \right],
\end{equation}
where $X^9_0$ commutes with all the matrices.
Then it is straightforward to see that equations~\eqref{AdS3SUSYcond}
are satisfied.  This solution describes a longitudinal fivebrane of
a (fuzzy) paraboloidal shape~\eqref{parab}.  The fivebrane wraps the entire
``$AdS_3\times S^3$ directions'' but it extends into the M-theory
direction through
the paraboloid equation.
\iftoomuchdetail
\begin{detail}%
(Note that as $\vartheta\rightarrow 0$, $X^a\sim
\sqrt{\vartheta}\rightarrow 0$ leaving
$X^9$ finite.)
\end{detail}%
\fi

\iftoomuchdetail
\begin{detail}%
The reader may have observed that equations~\eqref{AdS3SUSYcond} are just the
self-dual Yang-Mills equations, and are solved by the Banks-Casher
instanton\cite{bc}.  However, the remainder of the
equations~\eqref{AdS3SUSYcond} imply that $X^9$ is a specific SO(4)
generator, which cannot be realized within the Hilbert space of the
Banks-Casher instanton.  It therefore appears that the
Banks-Casher instanton cannot be generalized to give additional fully
supersymmetric solutions.
\end{detail}%
\fi

\section{THE 26 SUPERCHARGE MATRIX THEORY VACUA}

The \ppwave\ with 26 supercharges was presented in\cite{jm2}.
\iftoomuchdetail
\else
Supersymmetric vacua obey ($I,J,\dots=1\dots 7$, 
$I'',J'',\dots=1,2,3$, with the conventions\cite{jm2})
\fi
\begin{equation} \label{BPS26}
\begin{gathered}
\begin{aligned}
D_\tau X^I &= 0, I=1\dots 7, &
D_\tau X^8 &= \frac{\mu}{2} X^9, &
D_\tau X^9 &= -\frac{\mu}{2} X^8,
\end{aligned}
\\ \begin{aligned}
X^4 = X^5 = X^6 = X^7 &= 0, & \qquad
\com{X^{I''}}{X^{J''}} &= -i \frac{\mu}{R} \epsilon_{I''J''K''} X^{K''}, 
%I'',J'',K'' = 1,2,3.
\end{aligned} \\
\com{X^8}{X^{I''}} = 0 = \com{X^9}{X^{I''}}.
\end{gathered}
\end{equation}
That is, states preserving all the \supersymmetries\ 
\iftoomuchdetail
\begin{detail}%
preserved by $\proj_{4567}$
\end{detail}%
\fi
are fuzzy spheres at the origin of the
$(4,5,6,7)$-hyperplane, and orbit in the $(8,9)$-plane with
frequency~$\frac{\mu}{2}$%
\iftoomuchdetail
\begin{detail}
,
\begin{align} \label{26BPSrot}
X^8 &= r \sin \bigl(\tfrac{\mu}{2} \tau\bigr), &
X^9 &= r \cos \bigl(\tfrac{\mu}{2} \tau\bigr),
\end{align}
where the matrix $r$ must commute with $X^{I''}$, and therefore
specifies the radius of the circle traversed by each irreducible
component of the fuzzy sphere
\end{detail}%
\fi
.

\iftoomuchdetail
\begin{detail}
To verify that these solutions in fact preserve all 10
\supersymmetries, note that we no longer need $\hat{\Omega}_{4\dots7}$,
and that
\begin{equation}
%\begin{gathered}
\begin{aligned}
\hat{\Omega}_1 \proj_{8} 
   &= -\frac{\mu}{2} \Gamma^{23} \proj_{8}, \\
\hat{\Omega}_2 \proj_{8} 
   &=  \frac{\mu}{2} \Gamma^{13} \proj_{8}, \\
\hat{\Omega}_3 \proj_{8} 
   &= -\frac{\mu}{2} \Gamma^{12} \proj_{8},
\end{aligned} \qquad 
\begin{aligned}
\hat{\Omega}_8 \proj_{8} &=  \frac{\mu}{4} \Gamma^{9}  \proj_{8}, \\
\hat{\Omega}_9 \proj_{8} &= -\frac{\mu}{4} \Gamma^{8}  \proj_{8},
\end{aligned}
%\end{gathered}
\end{equation}
exactly as for the 8 \supersymmetries\ already considered.  Thus, the
conditions~\eqref{BPS26} are necessary and sufficient for preserving
all ten \supersymmetries, and we see that the most general fully
supersymmetric vacua are the rotating fuzzy spheres.  One easily
checks that these do satisfy the equations of motion.  It is
interesting that such states exist in the BMN
model\cite{bmn,dms}, but only preserve eight supercharges there.
\end{detail}%
\fi

Recall\cite{jm2} that the 26 supercharge \ppwave\ can be
compactified to a 26 supercharge IIA \ppwave.  Such a reduction occurs
along a Killing vector which rotates in the (8,9)-plane.
\iftoomuchdetail
\begin{detail}
Specifically, upon making the field redefinition
\begin{equation}
\begin{aligned}
X^8 &= X_{\text{IIA}}^9 \cos \frac{\mu}{2} \tau 
  + X_{\text{IIA}}^8 \sin \frac{\mu}{2} \tau, \\
X^9 &= -X_{\text{IIA}}^9 \sin \frac{\mu}{2} \tau 
  + X_{\text{IIA}}^8 \cos \frac{\mu}{2} \tau,
\end{aligned}
\end{equation}
$X_{\text{IIA}}^9$ becomes the compact M-theory direction, and
$X_{\text{IIA}}^8$ is the visible IIA 8-direction. 
Comparing to Eq.~\eqref{26BPSrot}, we see that
\end{detail}%
\else
Thus,
\fi
in the
IIA theory, the fully supersymmetric solutions~\eqref{BPS26} are
static%
\iftoomuchdetail
\begin{detail}
, at fixed $X_{\text{IIA}}^8$, and, in the Matrix String Theory, have
vanishing Wilson line ($X_{\text{IIA}}^9 \rightarrow A_\sigma$)%
\end{detail}%
\fi.
This is reminsicent of the maximally supersymmetric
\ppwave, for which\cite{dms} the orbiting fuzzy spheres broke the same
half 
of the (supernumerary) \supersymmetries\ as
were broken by the reduction to the IIA theory, yielding
fully (24 supercharge) supersymmetric IIA fuzzy
spheres located at any static value of $X_{\text{IIA}}^8$.  Here, the
solutions have identical form, but with two additional supercharges.

\section{IIB T-dual Matrix Theory Vacua} \label{sec:TIIB}

The T-dual, lifted to M-theory, of the
IIB maximally supersymmetric \ppwave\ is\cite{jm}
(the coordinates have been reindexed relative to the reference)
\begin{subequations} \label{IIBpp}
\begin{align}
ds^2 &= 2 dx^+ dx^- - \left[ 4 \mu^2 \sum_{I=1}^6 (x^I)^2 + 16 \mu^2
   (x^7)^2 \right] (dx^+)^2
   + \sum_{i=1}^9 (dx^i)^2, \\
F &= -4 \mu dx^+ \wedge dx^7 \wedge dx^8 \wedge dx^9
     +8 \mu dx^+ \wedge dx^5 \wedge dx^6 \wedge dx^7.
\end{align}
\end{subequations}
The linear SUSYs of the Matrix Theory
are parametrized by \hbox{$\Gamma^{5689}\epsilon_0 = \epsilon_0$}.

Fully supersymmetric vacua must be static, at
\hbox{$X^{1\dots4}=0$} and satisfy
\begin{subequations} \label{BPS}
\begin{gather}
%\begin{align}
%D_\tau X^i &= 0, & X^1=X^2=X^3=X^4=0,
%\end{align} \\
\label{longnotalg}
-2\frac{\mu}{R} X^7 + \frac{i}{2} \com{X^8}{X^9}
   - \frac{i}{2} \com{X^5}{X^6} = 0, \\
\com{X^6}{X^7} = 2 i \frac{\mu}{R} X^5, \quad
\com{X^7}{X^5} = 2 i \frac{\mu}{R} X^6, \\
\com{X^5}{X^8} = -\com{X^6}{X^9}, \quad
\com{X^6}{X^8} = \com{X^5}{X^9},
\end{gather}
\end{subequations}
\hbox{$\com{X^7}{X^8} = 0 = \com{X^7}{X^9}$}.
These equations are similar to those
for nonmaximally SUSic vacua of the BMN theory\cite{jhp}.
Solutions to equation~\eqref{BPS} satisfy the e.o.m.s.

For finite $N$, the only solutions are fuzzy ellipsoids,
(\hbox{$\com{J^a}{J^b}=i\epsilon_{abc} J^c$})
\begin{align} \label{fuzzy}
X^5 &= 2 \sqrt{2} \frac{\mu}{R} J^1, &
X^6 &= 2 \sqrt{2} \frac{\mu}{R} J^2, &
X^7 &= 2 \frac{\mu}{R} J^3.
\end{align}
Constant values of $X^{8,9}$, that
are diagonal and proportional to the identity in each irreducible
SU(2) block, give the positions of the fuzzy ellipsoids.

At strictly infinite $N$ there are many more solutions.
For example,
\begin{subequations} \label{ncplane}%
\begin{align}
X^{5,6} &= 2 \sqrt{2} \frac{\mu}{R} J^{1,2} \otimes \one, &
%X^6 &= 2 \sqrt{2} \frac{\mu}{R} J^2 \otimes \one, &
X^7 &= 2 \frac{\mu}{R} \left(J^3 - \frac{1}{8} \vartheta \one \right)
    \otimes \one, \\
X^8 &= \one \otimes \hat{x}^8, &
X^9 &= \one \otimes \hat{x}^9,
\end{align}
\end{subequations}%
where $\com{\hat{x}^8}{\hat{x}^9}=i \vartheta$ is a c-number.
That is, $X^{5,6,7}$ form a fuzzy ellipsoid and
$X^8$ and $X^9$ parameterize a noncommutative plane orthogonal to the
ellipsoid and translated, along $X^7$, by an amount 
\hbox{$-\frac{\mu}{4 R}\vartheta$} from
the origin.
This is the longitudinal M5-brane, formed as a stack of
M2-branes\cite{bfss,bss}, with one M2-brane blown up into a fuzzy
ellipsoid.
\iftoomuchdetail
\begin{detail}%
There is five-brane charge density because, since $X^{5,6}$ commute with
$X^{8,9}$, \hbox{$4! X^{[5} X^6 X^8 X^{9]} = 2 \com{X^5}{X^6} \com{X^8}{X^9}
\neq 0$}.
\end{detail}%
\fi
Thus, the
solution~\eqref{ncplane} describes a longitudinal M5-brane of
topology $\ZR^{1,3} \times S^2$.

One can also find a solution for every complex simple Lie
algebra of rank 2.  If the algebra is not simply-laced---{\em i.e.\/}\ except
for {$\mathfrak{su}(3)$}---the algebra gives two inequivalent
solutions.  Explicitly, if the two roots of the algebra are
\hbox{$\alpha_1 = (a_1,-b_1)$} and \hbox{$\alpha_2=(0,a_2)$}, with
Cartan subalgebra $h_1,h_2$, (\hbox{$\com{h_1}{\alpha_1}=a_1
  \alpha_1$}, \hbox{$\com{\alpha_2}{\alpha_2^\dagger}=a_2 h_2$},
{\em etc.\/}), then
\begin{equation} \label{gensoln1}
\begin{gathered}
\begin{aligned}
X^7 &= \frac{2 \mu}{a_1 R} h_1, & \qquad
X^5 &= \frac{2 \mu}{a_1 R} (e_1 + e_1^\dagger), & \qquad
X^6 &= \frac{2 \mu}{a_1 R i} (e_1 - e_1^\dagger),
\end{aligned} \\
\begin{aligned}
X^8 &= -2 i \sqrt{\frac{b_1}{a_1^2 a_2}} \frac{\mu}{R} (e_2 +
   e_2^\dagger), & \qquad
X^9 &= -2 \sqrt{\frac{b_1}{a_1^2 a_2}} \frac{\mu}{R} (e_2 - e_2^\dagger),
\end{aligned}
\end{gathered}
\end{equation}
is a solution.
Demanding hermitian matrices then restricts to a
particular noncompact realification of each algebra.
The resulting
realifications are ${\mathfrak{su}}(2,1)$, ${\mathfrak{sp}}(1,1)\cong
{\mathfrak{so}}(1,4)$, ${\mathfrak{sp}}(2,\ZR)\cong
{\mathfrak{so}}(2,3)$ and ${\mathfrak{g}}_{2(2)}$.  Two inequivalent
solutions are obtained from the last algebra.  As these are all
noncompact, their nontrivial unitary representations (hermitian
matrices) are all
infinite dimensional.

For example, explicit\cite{g1} ${\mathfrak{su}}(2,1)$ solutions can be written
in terms of three commuting sets of
annihilation operators $a_1,a_2,b$, and their hermitian conjugate
creation operators,
\begin{equation} \label{explSU21}
\begin{gathered}
\begin{aligned}
X^5 &= 2 \sqrt{\frac{2}{3}} \frac{\mu}{R}
   \left(a_1^\dagger a_2 + a_2^\dagger a_1 \right), \qquad &
X^6 &= -i 2 \sqrt{\frac{2}{3}} \frac{\mu}{R}
   \left(a_1^\dagger a_2 - a_2^\dagger a_1 \right),
\\
X^8 &= \frac{2}{\sqrt{3}} \frac{\mu}{R}
   \left( a_1 b + a_1^\dagger b^\dagger \right), &
X^9 &= -i \frac{2}{\sqrt{3}} \frac{\mu}{R}
   \left( a_1 b - a_1^\dagger b^\dagger \right),
\end{aligned} \\
X^7 = \frac{2}{3} \frac{\mu}{R}
   \left(a_1^\dagger a_1 - 2 a_2^\dagger a_2 - b b^\dagger \right).
\end{gathered}
\end{equation}
The quadratic Casimir of
${\mathfrak su}(2,1)$ takes the form
\begin{multline} \label{C2su21} \raisetag{2.5\baselineskip}
C_2^{{\mathfrak su}(2,1)} = 
-\frac{3}{4} \frac{R^2}{\mu^2} \left((X^8)^2 + (X^9)^2\right)
+\frac{3}{8} \frac{R^2}{\mu^2} \left( (X^5)^2 + (X^6)^2 + 2 (X^7)^2 \right)
\\
-\frac{9}{64} \frac{R^4}{\mu^4} \com{X^8}{X^9}^2
+\frac{9}{32} \frac{R^4}{\mu^4} \left( \com{X^5}{X^8}^2
    + \com{X^5}{X^9}^2 \right).
%= \frac{4}{3} \left( O_N + \frac{3}{2} \right)^2 - 3.
\end{multline}
%where \hbox{$O_N = a_1^\dagger a_1 + a_2^\dagger a_2 - b b^\dagger$}
%is the graded number operator, which is constant on irreducible
%representations.
The first line suggests a hyperboloidal interpretation to the solution, though
this needs refinement.

Compactifying the Matrix Theory along $X^8$ gives\cite{jmno}, at least for weak
string coupling, the
Green-Schwarz action for the IIB string in the \ppwave\ background.
The radius of compactification matches precisely.
This demonstration requires a field redefinition in the Matrix String
Theory that is the inverse of the coordinate
transformation that makes the isometry manifest\cite{jm} on the IIB side.
\iftoomuchdetail
\begin{detail}%
That is the subject of Sec.~\ref{sec:2+1}.
\end{detail}%
\else
Unfortunately, there is no space to give details here.
\fi

%%%%%%%%%%%%%%%%%%%%%%%%%%%%%%%%%%%%%%%%%%%%%%%%%%%%%%%%%%%%%
%                                                           %
% You may repeat \section{SECTION N-th HEADING TYPE HERE}   %
%                                                           %
% Do start a subsection or sub-subsection, do this:-        %
%                                                           %
%   \subsection{SUBSECTION HEADING TYPE HERE}               %
%                                                           %
%   \subsubsection{SUBSUBSECTION HEADING TYPE HERE}         %
%                                                           %
% instead of the above                                      %
%                                                           %
%%%%%%%%%%%%%%%%%%%%%%%%%%%%%%%%%%%%%%%%%%%%%%%%%%%%%%%%%%%%%

%\section{CONCLUSIONS}

%%%%%%%%%%%%%%%%%%%%%%%%%%%%%%%%%%%%%%%%%%%%%%%%%%%%%%%%%%%%%
% Doing Acknowledgement                                     %
%%%%%%%%%%%%%%%%%%%%%%%%%%%%%%%%%%%%%%%%%%%%%%%%%%%%%%%%%%%%%

\section*{Acknowledgments}

This work was supported in part by Department of Energy contract
\#DE-FG01-00ER45832% and NSF grant \#PHY-0071312%
.

%%%%%%%%%%%%%%%%%%%%%%%%%%%%%%%%%%%%%%%%%%%%%%%%%%%%%%%%%%%%%
% Doing Appendix(ices)                                      %
%%%%%%%%%%%%%%%%%%%%%%%%%%%%%%%%%%%%%%%%%%%%%%%%%%%%%%%%%%%%%

\appendix

%\section{HEADING FOR APPENDIX A}

%\renewcommand{\theequation}{A.\arabic{equation}}

\end{document}